\documentclass[preprint,aps]{revtex4}

\usepackage{graphicx}
\usepackage{dcolumn}
\usepackage{bm}
\usepackage{color}
\usepackage{subfigure}
\usepackage{epsfig}

\begin{document}

\preprint{}

\title{Polarization transfer in $^4$He$(\vec{e},e'\vec{p})$ and
  $^{16}$O$(\vec{e},e'\vec{p})$ \\ in a relativistic Glauber model}%

\author{P. Lava}
\author{J. Ryckebusch}
\author{B. Van Overmeire}
\affiliation{Department of Subatomic and Radiation Physics, Ghent
  University,
  Proeftuinstraat 86, B-9000 Gent, Belgium}
\author{S. Strauch}
\affiliation{Department of Physics, The George Washington University,
  Washington, D.C. 20052, USA}

\date{\today}

\begin{abstract}
Polarization-transfer components for $^4$He$(\vec{e},e'\vec{p})^3$H
and $^{16}$O$(\vec{e},e'\vec{p})^{15}$N are computed within the
relativistic multiple-scattering Glauber approximation (RMSGA). The
RMSGA framework adopts relativistic single-particle wave functions and
electron-nucleon couplings. The predictions closely match those of a
relativistic plane-wave model indicating the smallness of the
final-state interactions for polarization-transfer components. Also
short-range correlations play a modest role for the studied
observables, as long as small proton missing momenta are probed in
quasi-elastic kinematics.  The predictions with free and various
parameterizations for the medium-modified electromagnetic form factors
are compared to the world data.
\end{abstract}

\maketitle

\section{Introduction}

In conventional nuclear physics, nuclei are described in terms of
point-like protons and neutrons, interacting through the exchange of
mesons. It has been a long-standing and unresolved issue whether the
electromagnetic properties of bound nucleons differ from those of free
nucleons. Any sizable modification would have a severe impact on the
interpretation of e.g. the EMC effect \cite{miller03}.  Inclusive
$A(e,e')$ data, including their separated longitudinal and transverse
cross sections, are rather inconclusive with respect to the allowed
ranges for medium modifications. Indeed, a recent re-analysis of the
longitudinal inclusive $^{4}$He$(e,e')$ response, implementing
two-body effects in the nuclear charge operator and realistic wave
functions, finds the data consistent with the state-of-the-art
calculations when using free-nucleon electromagnetic form factors
\cite{carlson2003}.  To the contrary, an alternate recent
re-evaluation of the Coulomb sum rule (CSR) concentrating on heavier
nuclei, discerns it considerably quenched for $A \geq 40$, thereby not
excluding sizable medium modifications for the electric form factor
$G_E(Q^2)$~\cite{morgenstern2003}. A $y$-scaling analysis of the
inclusive $A(e,e')$ data \cite{sick85}, on the other hand, indicates
that the medium effects on the magnetic form factor $G_M(Q^2)$ are
smaller than $3\%$ for $Q^2 \ge 1$~(GeV/c)$^2$. At lower values of the
four-momentum transfer $Q^2$, a considerably improved description of
the separated longitudinal and transverse $A(e,e')$ responses for
$^{12}$C and $^{40}$Ca was reached after including in-medium
$G_E(Q^2)$ and $ G_M(Q^2)$ form factors as computed in the
Nambu-Jona-Lasinio model \cite{henley89}.  This model is thought to
provide a reasonable description of the dynamical breaking of chiral
symmetry at nuclear-physics' scales.

Exclusive $A(e,e'p)$ processes have been put forward as more
discriminative than inclusive $A(e,e')$ when it comes to investigating
specific aspects of nuclei, and in particular, the possible
modifications of the electromagnetic properties attributed to the
presence of a medium. Finding signatures of those medium
modifications, however, requires an excellent control over all those
ingredients of the $A(e,e'p)$ reaction process that are directly
related to the presence of a nuclear medium.  They include
medium-related effects, such as final-state interactions (FSI),
meson-exchange currents (MEC) and isobar currents (IC).  We wish to
stress that in principle there is a clear distinction between FSI, MEC
and IC effects and those dubbed ``medium modifications''.  Indeed, the
latter refer to medium-driven changes in the internal quark-gluon
structure of nucleons.  Unfortunately, at the level of the
$A(\vec{e},e'\vec{p})$ observables, no formal distinction can be
made between FSI, MEC and IC effects on one hand and possible medium
modifications.

In the eighties, it was suggested that the ratio of the transverse (T)
to the longitudinal (L) response in exclusive $A(e,e'p)$ may provide a
handle on the medium modifications of the nucleon's electromagnetic
properties \cite{gerard86,reffay88}.  The longitudinal-transverse
$A(e,e'p)$ separations suggested substantial deviations from the naive
(i.e., plane-wave impulse approximation (PWIA)) predictions for the
T/L ratio. The data for medium-heavy nuclei like $^{12}$C and
$^{40}$Ca, however, could be satisfactorily explained after
implementing FSI mechanisms \cite{ryckebusch89}, thereby adopting
free-nucleon electromagnetic form factors.  For the $^4$He nucleus,
charge-exchange processes turned out to be of great importance to
explain the measured T/L ratios \cite{buballa91,warmann91,ducret93}.
The above-mentioned findings indicate that medium modifications of the
electromagnetic form factors are apparently modest and support the
picture that despite their substructure, nucleons are rather robust
objects.

For a long time, Rosenbluth separations in elastic electron-proton
scattering were the sole source of information about free-proton
electromagnetic form factors.  Continuing efforts to improve on the
quality of electron beams and hadron detection, however, made precise
measurements of polarization degrees-of-freedom a viable option to
address issues in hadronic physics with the aid of the electromagnetic
probe. In polarized electron free-proton scattering $ \vec{e} (E_e) + p  
 \longrightarrow e' (E_{e'} ) + \vec{p}$, the ratio of the
electric ($G_E (Q^2=-q^{\mu}q_{\mu})$) to the magnetic ($G_M (Q^2)$)
Sachs form factors, can be extracted from \cite{arnold81}
\begin{equation}
\frac{G_E (Q^2)}{G_M (Q^2) } = -\frac{P'_x}{P'_z} \; \frac{E_e +
E_{e'}}{2M_p}\tan\left(\frac{\theta_e}{2}\right) \; .
\end{equation}
Here, $q^{\mu}$ is the four-momentum transfer, $P'_{x}$ and $P'_{z}$
is the transferred polarization in the direction perpendicular to and
parallel with the three-momentum transfer, and $\theta_e$ the electron
scattering angle.

Of all observables accessible in $A(e,e'p) $, the transferred
polarization components $P'_i$ have been recognized as the ones
with the weakest sensitivity to FSI, MEC and IC distortions
\cite{laget94,caballero98,udias99,udias2000,kelly99,kelly299}.
Therefore, polarization-transfer components have been put forward as a
tool to examine the magnitude of the in-medium electromagnetic form
factors.  Hereby, one adopts the philosophy that the in-medium (or,
off-shell) electron-proton vertex $\Gamma ^{\mu}$ has the same Lorentz
structure as the free-proton one. This is the so-called impulse
approximation (IA) which has been successfully applied in a vast
number of calculations. Investigations search for anomalous behavior,
which manifests itself as a deviation between up-to-date calculations
and well-controlled observables obtained in optimized kinematic
conditions.  Possible anomalous behavior of this kind may
subsequently be interpreted as an indication for a medium effect.  The
described procedure is a pragmatic one and may be subject to
criticism, particularly in view of the ambiguities with respect to
describing the off-shell $\Gamma ^{\mu}$ vertex \cite{naus90}.

Recently, $(\vec{e},e'\vec{p})$ measurements for the target nuclei
$^{16}$O \cite{malov00} and $^4$He \cite{dieterich01,strauch03} have
been reported. The $^{16}$O$(\vec{e},e'\vec{p})$ measurements have
been confronted to various non-relativistic and relativistic
calculations
\cite{ryckebusch99,meucci01,radici03,kazemi03,martinez04}.  All these
calculations utilize an optical potential to incorporate the FSI. The
calculations of Ref.~\cite{kazemi03} indicate that two-nucleon
currents like MEC and IC affect the polarization-transfer components
in $^{16}$O to less than 5\% provided that missing momenta below
$200$~MeV/c are probed. The non-relativistic calculations of
Ref.~\cite{ryckebusch99} attributed somewhat larger corrections to the
two-nucleon currents, in particular for proton knockout from the
$p3/2$ and $s1/2$ shells.  All calculations, however, predict similar
trends for the MEC and IC corrections on the polarization-transfer
components.  One major finding is that their effect dwindles with
increasing $Q^2$ and decreasing missing momentum. Relativistic effects
on the transferred polarizations $P'_x$ and $P'_z$ have been
investigated in Refs.~\cite{meucci01,martinez04} and are discerned at
the few percent level as long as the probed missing momentum remains
relatively small ($p_m \leq 200$ MeV/c).  These studies also indicated
that at higher missing momenta the uncertainties stemming from
off-shell ambiguities are larger than the overall impact of the
relativistic effects.  Apparently, all theoretical investigations
indicate that when probing low missing momenta in quasi-elastic
kinematics, the effect on the polarization-transfer components of
typical medium-related complications like MEC, IC and off-shell
ambiguities can be kept under reasonable control.

In Ref.~\cite{strauch03} the Jefferson Laboratory (JLAB)
$^{4}$He$(\vec{e},e'\vec{p})$ data, which cover the range $0.5 \le Q^2
\le 2.6~$(GeV/c)$^2$, are compared to the state-of-the-art relativistic
distorted-wave impulse approximation (RDWIA) calculations of Ud\'{i}as
\textit{et al.}  \cite{Udias93}.  This model provided a better overall
description of the data when implementing medium-modified
electromagnetic form factors as predicted in the Quark-Meson Coupling
(QMC) model \cite{guichon88,lu98,lu99}. At JLAB, exclusive $A(e,e'p)$
studies are conducted in a kinematic regime which may outreach the
range of applicability of optical-potential approaches for describing
FSI mechanisms. Indeed, given the highly inelastic and diffractive
nature of proton-nucleon scattering at proton lab momenta exceeding 1
GeV/c, the use of optical potentials for modeling FSI seems rather
unnatural.  For example, for the $Q^2$=2.6~(GeV/c)$^2$ case, the
$^{4}$He$(\vec{e},e'\vec{p})$ data of Ref.~\cite{strauch03} are
compared to RDWIA calculations with extrapolated optical potentials. 

At higher energies, Glauber multiple-scattering theory provides a more
natural and economical description of FSI mechanisms
\cite{glauber70,Wallace75,Yennie71}. A typical Glauber model is based
on the eikonal approximation and the assumption of consecutive
cumulative scattering of a fast proton on a composite target of $A-1$
``frozen'' nucleon scatterers. Recently, we developed a relativistic
version and dubbed it as the relativistic multiple-scattering Glauber 
approximation (RMSGA) \cite{debr02npa,debr02plb,ryckebusch03}.  The
RMSGA heavily draws on ingredients of standard RDWIA $A(e,e'p)$
approaches.  For example, the assumptions made with respect to the
construction of the bound-state wave functions and electromagnetic
couplings are identical in RDWIA and RMSGA.  The sole difference
concerns the construction of the scattering wave function. In RDWIA
one adopts the philosophy that optical potentials parameterizing the
FSI mechanisms in elastic $A(p,p)A$ processes, can
also be utilized to model the impact of the proton's distortions in
$A(e,e'p)$ reactions. In a Glauber framework, on the other
hand, the effects of FSI are computed directly from nucleon-nucleon
scattering data. Despite the dissimilar assumptions underlying the
treatment of FSI, it has been recently shown that the RMSGA and RDWIA
predictions for the nuclear transparencies extracted from $A(e,e'p)$
are alike in an intermediate proton kinetic-energy range where both
optical-potential and Glauber approaches are judged to be applicable
\cite{lava03}.  The RMSGA $A(e,e'p)$ model has a number of virtues
including the fact that it is unfactorized, which means that our cross
section is not written in terms of the product of an off-shell
electron-proton cross section and a distorted missing momentum distribution. 
In addition, our implementation adopts the
full-fledged multiple-scattering version of the Glauber approach and
describes each nucleon scattering center in the residual nucleus with
its particular single-particle wave function. Thereby, we avoid a
frequently adopted averaging approximation which allows introducing
the nuclear density.

In this paper, RMSGA predictions for the polarization-transfer
components in $^4$He and $^{16}$O will be presented and compared to
the world data.  The numerical calculations are performed with both
free and medium-modified electromagnetic form factors.  For the latter
we use the predictions of the QMC model
\cite{guichon88,lu98,lu99} and of a modified Skyrme model
\cite{yakshiev02,meisner03}. It is the purpose of this paper to
address the questions whether a Glauber approach can adequately
describe the $(\vec{e},e'\vec{p})$ polarization-transfer components
and to what extent its predictions  differ from typical
distorted-wave (or, optical-potential) approaches. 

The outline of this paper is as follows. In Sect.~\ref{sec:rmsga} the
basic features of the RMSGA formalism are sketched.
Sect.~\ref{sec:ffactor} presents predictions for the medium-modified
electromagnetic form factors from some specific nucleon models and
outlines how these form factors are implemented in the calculation of
the polarization-transfer components. Sect.~\ref{sec:results}
presents our numerical results. We summarize our findings and state
our conclusions in Sect.~\ref{sec:conc}.
  
\section{Formalism}

In this section, we first review the basic ingredients which enter the
RMSGA formalism \cite{ryckebusch03}. Next, the method of implementing
medium-modified electromagnetic form factors is outlined.

\subsection{RMSGA model}
\label{sec:rmsga}

Adopting the IA and the independent-nucleon
picture, the basic quantity to be computed in a relativistic approach
to $A(e,e'p)$ is the transition matrix element
\begin{equation}
\langle J^{\mu}
\rangle = \int
d\vec{r} \; \overline{\phi}_F(\vec{r})\hat{J}^{\mu}(\; \vec{r} \;) e^{i\vec{q}.\vec{
r}}\phi_{\alpha}( \; \vec{r} \; ) \; ,
\label{eq.:relcurrent}
\end{equation}
where $\phi_{\alpha}$ and $\phi_F$ are the relativistic bound-state and
scattering wave functions. Further, $\hat{J}^{\mu}$ is the
relativistic one-body current operator modeling the coupling between
the virtual photon and a nucleon embedded in the medium. The
relativistic bound-state wave functions are obtained within the
Hartree approximation to the $\sigma - \omega$ model \cite{serot86}.
As discussed by Walecka \cite{walecka74}, the quantum-field theory can
be approximated by replacing the meson field operators with their
expectation values. The resulting eigenvalue equations of the
relativistic mean-field theory can be solved exactly.  The
corresponding bound-state wave functions $\phi_{\alpha}$ are four-spinors and
can be formally written as
\begin{equation}
\phi_{\alpha}(\; \vec{r}, \vec{\sigma} \; )=\left(
                \begin{array}{@{\hspace{0pt}}c@{\hspace{0pt}}} \frac
                {i G_{n_{\alpha} \kappa_{\alpha}}(r)} {r} \mathcal{Y}
                _{\kappa_{\alpha} m_{\alpha}} (\Omega_r,\vec{\sigma})
                \\ \frac {- F_{n_{\alpha} \kappa_{\alpha}}(r)} {r}
                \mathcal{Y} _{- \kappa_{\alpha}
                m_{\alpha}}(\Omega_r,\vec{\sigma}) \\
             \end{array} \right) \, ,
\label{bwf}
\end{equation}
with $\mathcal{Y} _ {\kappa _{\alpha} m_{\alpha}}(\Omega_r,
\vec{\sigma})$ the usual spin spherical harmonics.  In a
high-resolution and exclusive $A(e,e'p)$ experiment, the angular
momentum of the state in which the $A-1$ residual nucleus is left,
determines the quantum numbers $\alpha \equiv (n_{\alpha},\kappa
_{\alpha}, m_{\alpha})$. In determining the bound-state wave
functions, all results contained in this work use the W1
parametrization \cite{furnstahl97} for the different field strengths.
Further, we adopt the Coulomb gauge and the current operator in its
$CC2$ form \cite{Forest1983}
\begin{equation}
\label{eq.:currentoperator}
J^{\mu}(\vec{r})= F_{1}^{p}(Q^2) \gamma ^{\mu} + F_{2}^{p} (Q^2) i
  \frac {\kappa _p} {2 m_p } \sigma ^{\mu \nu} q _{\nu} \; .
\end{equation}
In computing the matrix elements, the $q^{\mu}$ is evaluated in the
laboratory frame and the energy transfer is based upon
electron-scattering kinematics.

We now turn to the question of how to determine a relativistic
scattering wave function for the emitted proton.  Traditionally, the
Glauber approach relies on a number of assumptions.  First, the use of
the eikonal approximation, and, further, the so-called frozen
approximation.  The latter allows one to formulate a full-fledged
multiple-scattering theory for the emission of a ``fast'' proton from
a composite system consisting of $A-1$ frozen nucleons.  In
Ref.~\cite{ryckebusch03}, a relativistic and unfactorized formulation
of Glauber multiple scattering theory has been outlined.  In this
approach, termed RMSGA, the scattering wave function in the matrix
element of Eq.~ (\ref{eq.:relcurrent}) takes on the
following form
\begin{equation}
\label{eq.:transition}
\phi_F({\vec{r}}) \equiv \phi _{p_F, \; s_F}(\vec{r}) \; \mathcal{G}
(\vec{b},z),
\end{equation}
 where $\phi_{p_F, \; s_F}$ is a relativistic plane wave.  The impact
 of the FSI mechanisms on the scattering wave function is contained in
 the scalar Dirac-Glauber phase $\mathcal{G}(\vec{b},z)$
\begin{equation}
\label{eq.:glauberphase}
\mathcal{G}(\vec{b},z)= \prod_{\alpha \neq \alpha_{1}} \biggl[ 1-\int
d\vec{r}~'|\phi_{\alpha}(\vec{r}~')|^2 \theta(z'-z)\Gamma(\vec{b}
-\vec{b}') \biggr] ,
\end{equation}
where the product over $\alpha (n, \kappa, m)$ extends over all
occupied single-particle states in the target nucleus, not including
the one (here denoted as $\alpha_{1}$) from which the proton is
ejected.  The profile function for $pN$ scattering is defined in the
standard manner
\begin{equation}
\label{eq.:profile}
\Gamma(\vec{b})=
\frac {\sigma_{pN}^{tot} (1-i\epsilon_{pN})} 
{4\pi\beta^2_{pN}} \exp(\frac{-b^2}{2\beta^2_{pN}})\;.
\end{equation}
The parameters $\sigma_{pN}^{tot}$, $\beta_{pN}$ and $\epsilon_{pN}$
depend on the proton energy and fitted values to the $pN$ data can be
found in Ref.~\cite{pdg}.

 The Dirac-Glauber phase $\mathcal{G}(\vec{b},z)$ of
 Eq.~(\ref{eq.:glauberphase}) can be cast in the following form
\begin{widetext}
\begin{eqnarray}
\mathcal{G}(\vec{b},z)  = \prod _{\alpha (n,\kappa,m) \ne {\alpha_{1}}
  (n_1, \kappa _1, m_1) }  
\Biggl\{1 - 
\frac{\sigma^{tot}_{pN}( 1 - i \epsilon _{pN})} {4\pi \beta _{pN}^2} 
\int _{0} ^{\infty} b'db' \int _{- \infty}
^{+ \infty} dz' \theta (z' -z) 
\nonumber \\
 \times \Biggl( \biggl[ {\frac{G_{n \kappa} \left( r'(b',z') \right)}
{r'(b',z')} }
\mathcal{Y} _{\kappa m} (\Omega ' , \vec{\sigma} )
\biggr] ^2  + \biggl[ {\frac{ F_{n \kappa} \left(
r'(b',z') \right) }
{r'(b',z')}}
\mathcal{Y} _{- \kappa m} (\Omega ', \vec{\sigma} ) \biggr]  ^2 \Biggr)  
\nonumber \\  \times
\exp \left[ -\frac{(b - b')^2 }{2 \beta _{pN} ^2} \right] 
\int _{0} ^{2 \pi} d \phi_{b'}
\exp \biggl[ \frac{-b b'}{\beta_{pN}^2}2 {\sin}^2 \left( \frac {\phi_{b} -\phi_{
b'}} {2} \right) \biggr]
\Biggr\} \; .
\label{eq.:diracglauberphase}
\end{eqnarray} 
\end{widetext}
For numerical reasons, the $z$ axis is chosen along the asymptotic
direction of the ejectile.  It is noteworthy that when computing the
Dirac-Glauber phase $\mathcal{G}(\vec{b},z)$ each of the residual
nucleons behaves as a ``frozen'' scattering center with its unique
relativistic wave function, which has an upper ($G(r)$) and lower
($F(r)$) component. Cylindrical symmetry about the axis defined by the
ejectile's asymptotic momentum makes the Dirac-Glauber phase to depend
on the two independent variables $(b,z)$.  Hereby, $b=~\mid\vec{b}\mid
$, where $\vec{b}$ is orthogonal to the ejectile's direction.  The
expression (\ref{eq.:diracglauberphase}) includes the cumulative
effect of free passage of the hit proton, single-scattering,
double-scattering, up to and including ($A-1$)-fold scattering. Often,
the product over all scattering centers $\prod_{\alpha \neq \alpha_1}$
is approximated by a sum which is cut at some order in the
multiple-scattering series.  At the expense of a great numerical cost,
we compute the expression (\ref{eq.:diracglauberphase}) rigorously.

\begin{figure*}
\begin{center}
\mbox{\subfigure{\epsfig{figure=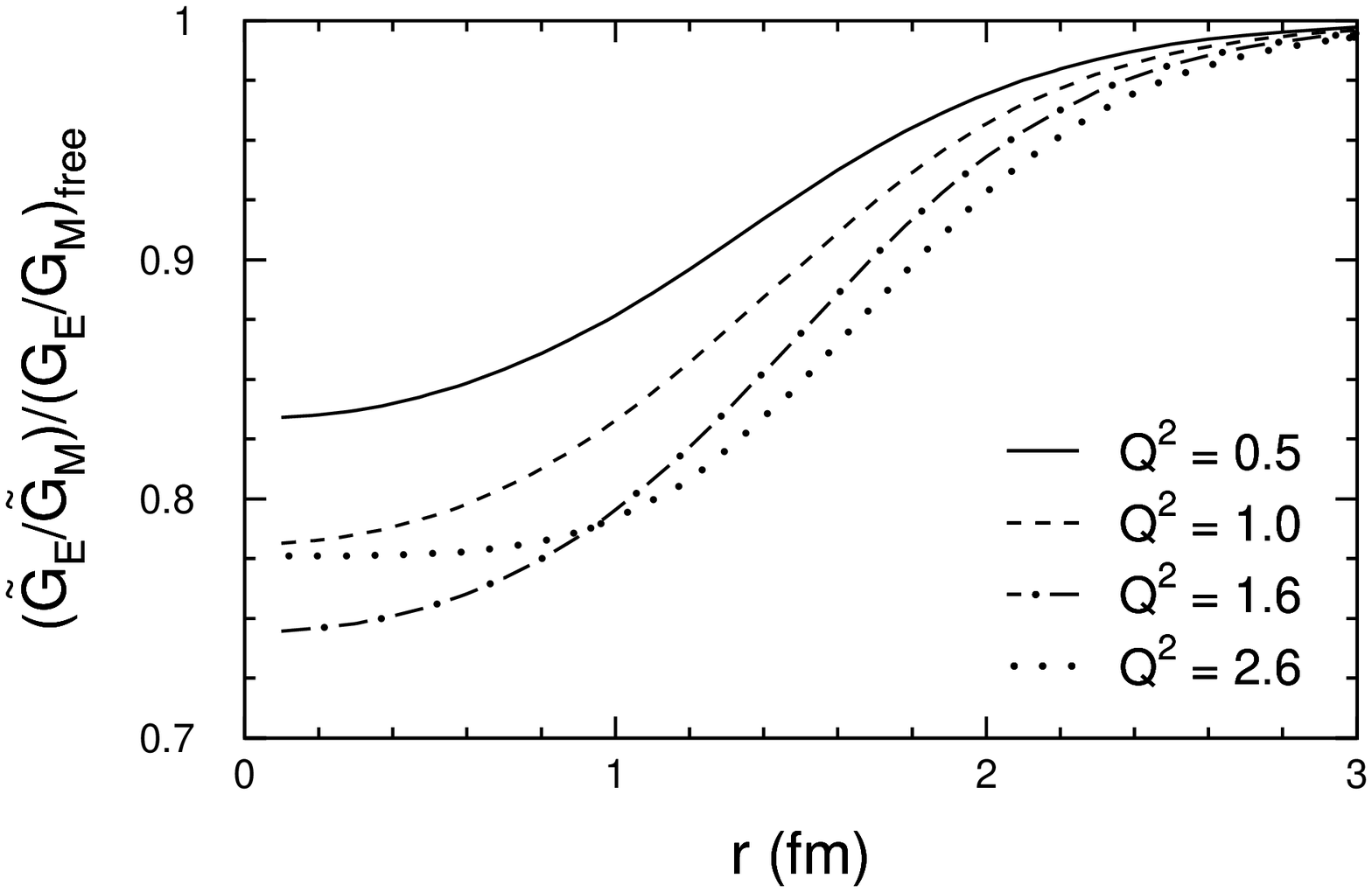,width=0.5\textwidth}}\quad
  \subfigure{\epsfig{figure=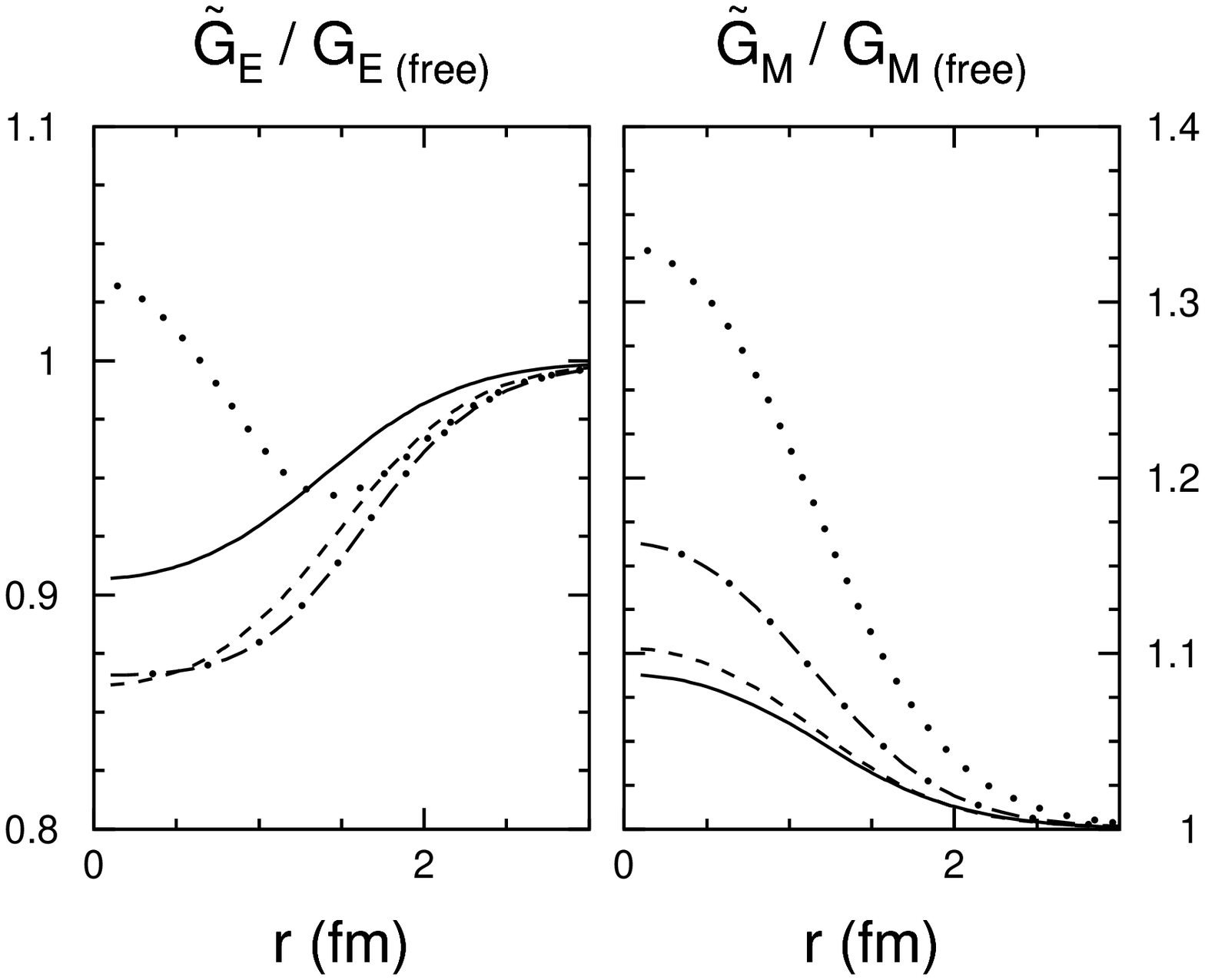,width=0.5\textwidth}}}
\end{center}
\caption{QMC predictions \cite{lu99} for the radial dependence of $G_E$, $G_M$ and
  $G_E/G_M$ in $^4$He at four different values of $Q^2$
  (GeV/c)$^2$. The bag radius was taken to be $0.8$~fm.}
\label{fig.:gegm}
\end{figure*}

\subsection{Electromagnetic form factors}
\label{sec:ffactor}

\begin{figure}
\begin{center}
\includegraphics[width=0.5\textwidth]{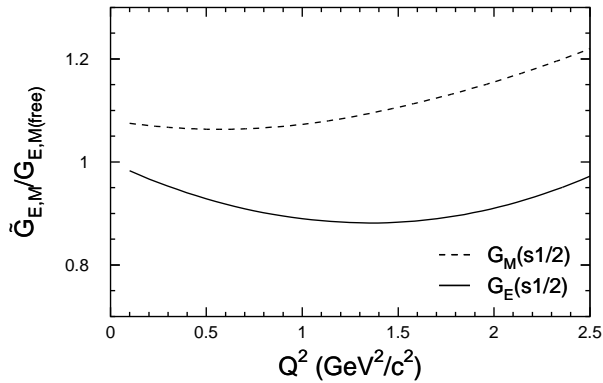}
\end{center}
\caption{The $Q^2$-dependence of the ratio of the in-medium to
  free electric and magnetic form factors for the proton in $^4$He
  according to the QMC model with a bag radius of 0.8~fm \cite{lu99}.}
\label{fig.:gint}
\end{figure}

In the QMC model \cite{guichon88,lu98,lu99}, the scalar ($\sigma$) and
vector ($\omega$) fields, carrying the forces between nucleons in
Quantum Hadrodynamics \cite{walecka74,serot86}, couple directly to the
quarks within the nucleon. As a result, the intrinsic properties of a
bound nucleon are affected by the presence of a medium. In the
QMC framework, the nucleon is described in terms of the MIT bag model
with almost massless and relativistic point-like quarks. For
the $A(\vec{e},e'\vec{p})$ results presented below, we use the QMC
predictions corresponding to a bag radius of $0.8$~fm.  In the QMC
model, the electric and magnetic form factors attain a dependence on
the total baryon density~: $ G_{E,M}(Q^2) \rightarrow
G_{E,M}(\rho_B(\vec{r}),Q^2)$.  In a mean-field model, the total baryon
density $\rho_B(\vec{r})$ is defined according to
\begin{equation}
\label{eq.:rho}
\rho_B(\vec{r}) = \sum_{\alpha} \int
d \vec{\sigma}(\phi_{\alpha}(\vec{r},\vec{\sigma}))^
{\dagger}(\phi_{\alpha}(\vec{r},\vec{\sigma})).
\end{equation}
The magnitude of the free form factors is not so well described within
the QMC model.  Therefore, we retain only the prediction for its density
dependence and scale the free form factor with the ratio of the QMC
form factors at a given density, to the ones at vanishing baryon
density
\begin{equation}
\widetilde{G}^{QMC}_{E,M}(\rho_B(\vec{r}),Q^2) = G_{E,M}(Q^2)
\frac{G_{E,M}^{QMC}(\rho_B(\vec{r}),Q^2)}
{G_{E,M}^{QMC}(\rho_B(\vec{r})=0,Q^2)} \; .
\label{eq:ffrescaled}
\end{equation}
In Fig.~\ref{fig.:gegm} the QMC predictions for the radial dependence
of $\widetilde{G}_E$, $\widetilde{G}_M$ and their ratio in $^4$He are
displayed at four different values of $Q^2$.  Thereby, we have plotted
the renormalized quantities as defined in
Eq.~(\ref{eq:ffrescaled}). The magnitude of medium modifications grows
with $Q^2$. As suggested by Kelly in Ref.~\cite{kelly299}, the $Q^2$
dependence of the above ratios for a particular single-particle state
can be estimated in the local density approximation in terms of the
following density convolution
\begin{equation}
\label{eq.:lda}
\widetilde{G}_{E,M}^{QMC}(\alpha_1,Q^2) = \frac {\int
  \widetilde{G}_{E,M}^{QMC} (\rho _B (\vec{r}),Q^2) \; \rho_{\alpha_1}
  (\vec{r}) \; d\vec{r}} {\int \rho_{\alpha_1} (\vec{r}) \; d\vec{r}} \;
  .
\end{equation}
Here, $\rho_{\alpha_1}(\vec{r})$ is the square of the $<A-1 \mid A>$
overlap wave function.  In a naive independent-particle picture the
overlap wave function corresponds with the single-particle wave
function of the state from which the proton is ejected.  Fig.~\ref{fig.:gint} displays
$\widetilde{G}_{E,M}^{QMC}(s1/2,Q^2)$ for a proton in $^4$He. At $Q^2
\geq 1.5$ (GeV/c)$^2$, the averaged medium magnetic form factor is
10\% larger than the free one.
It has been pointed out that modifying the bag radius can considerably
reduce the overall magnitude of the medium effects \cite{thomas98}. A
recent calculation in the Chiral Quark Soliton model resulted in
predictions for the electromagnetic form factors of bound protons
which show the same trends as the QMC model \cite{smith04}.

\begin{figure}
\begin{center}
\includegraphics[width=0.5\textwidth]{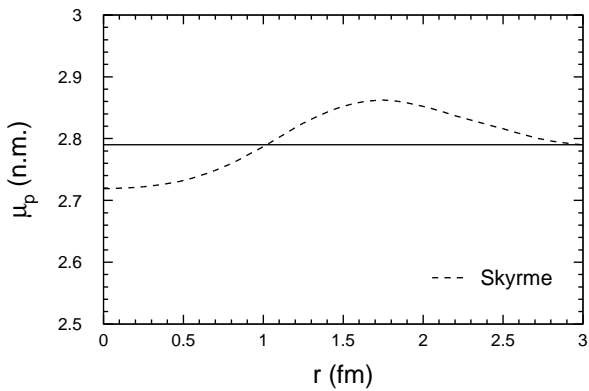}
\end{center}
\caption{The radial dependence of the proton
 magnetic moment in $^4$He according to the Skyrme model of Ref.~\cite{meisner03}. The
 free case corresponds with $\mu_p \mbox{=} 2.79$ n.m.}
\label{fig.:protmag}
\end{figure}

Recently, Yakhshiev et al.~\cite{yakshiev02,meisner03} addressed the
issue of in-medium electromagnetic form factors in the framework of a
modified Skyrme model. This model provides a fair description of
nucleon properties in free space and adopts degrees of freedom
directly related to the spontaneous chiral symmetry breaking of
QCD. In contrast to most constituent quark models, the pion-cloud
contribution is naturally taken into account. As a result, the
influence of the nuclear medium and the nucleon's response to it, is
predicted to be very probe dependent.  Beyond $Q^2 =0.6$~(GeV/c)$^2$,
vector mesons and boost effects are deemed to come into play, and the
Skyrme model is no longer considered realistic. In the Skyrme model,
the proton magnetic moment gains an additional radial dependence
dictated by the density of the nucleus.  Whereas $G_E(Q^2)$ remains
unaffected, its magnetic counterpart takes on the form
\begin{equation}
G_M(Q^2,r) = \mu_p(r)G_E(Q^2).
\end{equation}
In Fig.~\ref{fig.:protmag} the medium proton magnetic
moment is displayed as a function of the distance to the center of
the $^4$He nucleus. In the interior of $^4$He, the magnetic form
factor is mildly suppressed, whereas a modest increment is observed in
the surface area.

When including medium modifications in the $A(\vec{e},e'\vec{p})$
calculations, the electromagnetic current operator of
Eq.~(\ref{eq.:currentoperator}) is modified according to
\begin{equation}
\label{eq.:currentoperatordens}
J^{\mu}(\vec{r})= \widetilde{F}_{1}^{p}(\rho _B (\vec{r}) ,Q^2) \gamma ^{\mu} 
+ \widetilde{F}_{2}^{p} (\rho_B (\vec{r}), Q^2) i
  \frac {\kappa _p} {2 m_p } \sigma ^{\mu \nu} q _{\nu} \;.
\end{equation}
The density-dependent Dirac and Pauli form factors are related to the
$\widetilde{G}_E^{QMC}(\rho_B (\vec{r}) , Q^2)$ and
$\widetilde{G}_M^{QMC} (\rho_B(\vec{r}) , Q^2)$ of
Eq.~(\ref{eq:ffrescaled}) in the standard fashion.  The
medium-modified form factors $\widetilde{F}_{1,2}^{p}$ in
Eq.~(\ref{eq.:currentoperatordens}) depend upon the total density in the neighborhood
of the nucleon that absorbs the virtual photon.


\begin{figure}
\begin{center}
\includegraphics[width=0.4\textwidth]{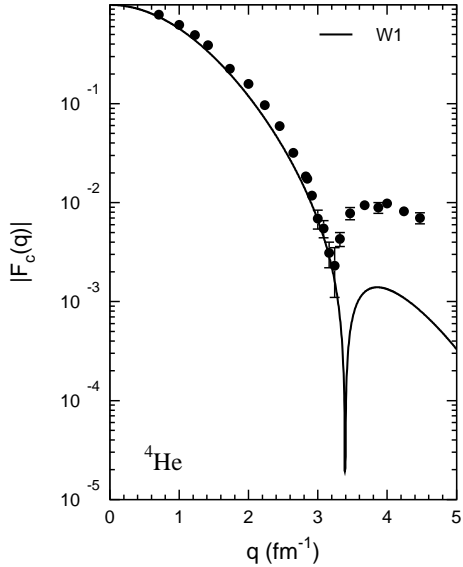}
\end{center}
\caption{The charge form factor of $^4$He, obtained within the W1
parametrization of Ref.~\cite{furnstahl97}. The data are from
\cite{frosch67} and \cite{mccarthy77}.}
\label{fig.:chargeform}
\end{figure}

\section{Results}
\label{sec:results}

All $^4$He$(\vec{e},e'\vec{p})$ and $^{16}$O$(\vec{e},e'\vec{p})$
calculations reported in this section are performed in quasi-elastic
kinematics and adopt kinematical conditions which allow a direct
comparison with the available data from
Refs.~\cite{dieterich01,strauch03, malov00}. For the $^4$He nucleus,
the polarization-transfer measurements have been performed in parallel
kinematics.

Throughout this section, we adopt a dipole parameterization for the
free-nucleon form factors. This choice may appear doubtful as improved
fits implementing the new $p(\vec{e},e')\vec{p}$ data are readily
available \cite{budd03}.  For the present purposes, however, a dipole
parameterization is adequate. Indeed, the
$^{16}$O$(\vec{e},e'\vec{p})$ data are restricted to $Q^2 \mbox{=}
0.8$~(GeV/c)$^2$, where deviations between the dipole and more
sophisticated parameterizations are minor.  The $^4$He
polarization-transfer results, on the other hand, are commonly expressed
in terms of a double ratio $R$ \begin{equation} R =
\frac{(P'_x/P'_z)_{^4He}}{(P'_x/P'_z)_{^1H}} \; ,
\end{equation}
which is almost independent of the used parameterization for the form
factors, as long as identical ones are used for $^4$He and $^1$H. In
order not to obscure the result by small kinematical differences
between the individual $^1$H and $^4$He measurements, data and
calculations are often shown in terms of a double ratio with the RPWIA
result as baseline.

At present, realistic relativistic wave functions for the $^4$He
ground state are not available.  Wave functions based on a
relativistic mean-field approach emerge as the only alternative when
embarking on fully relativistic $A(e,e'p)$ calculations.  At first
sight, an independent-particle approximation for describing the
four-nucleon system may appear as a venture into dangerous territory.
As can be appreciated from Fig.~\ref{fig.:chargeform}, however, a fair
description of the low-momentum part of the charge form factor for the
$^4$He nucleus is obtained with the ``W1'' parameterization used
throughout this work. The deviation between the computed and measured
charge form factor $F_c$ at high momentum transfer can be partly
attributed to large two-body charge contributions \cite{carlson1998},
which are neglected for the curve displayed in
Fig.~\ref{fig.:chargeform}.

\begin{figure}
\begin{center}
\includegraphics[width=0.5\textwidth]{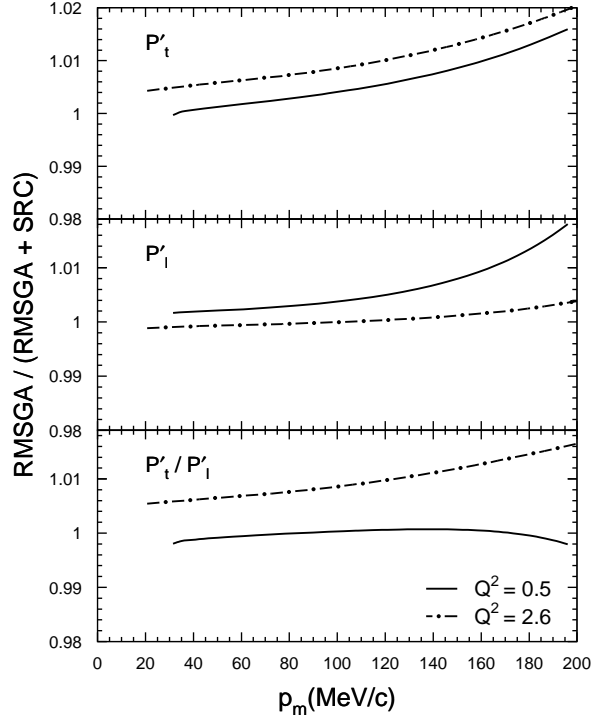}
\end{center}
\caption{Relative effect of short-range correlations on the
  polarization-transfer components and their ratio. The solid
  (dot-dashed) curves refer to $^4$He$(\vec{e},e'\vec{p})$ in
  quasi-elastic and parallel kinematics for $Q^2 \mbox{=}
  0.5~(2.6)$~(GeV/c)$^2$. The RMSGA + SRC results implement the effect
  of SRC according to the prescription of Eq.~(\ref{eq:srcglaub}).}
\label{fig.:src}
\end{figure}

\begin{figure*}
\begin{center}
\mbox{\subfigure{\epsfig{figure=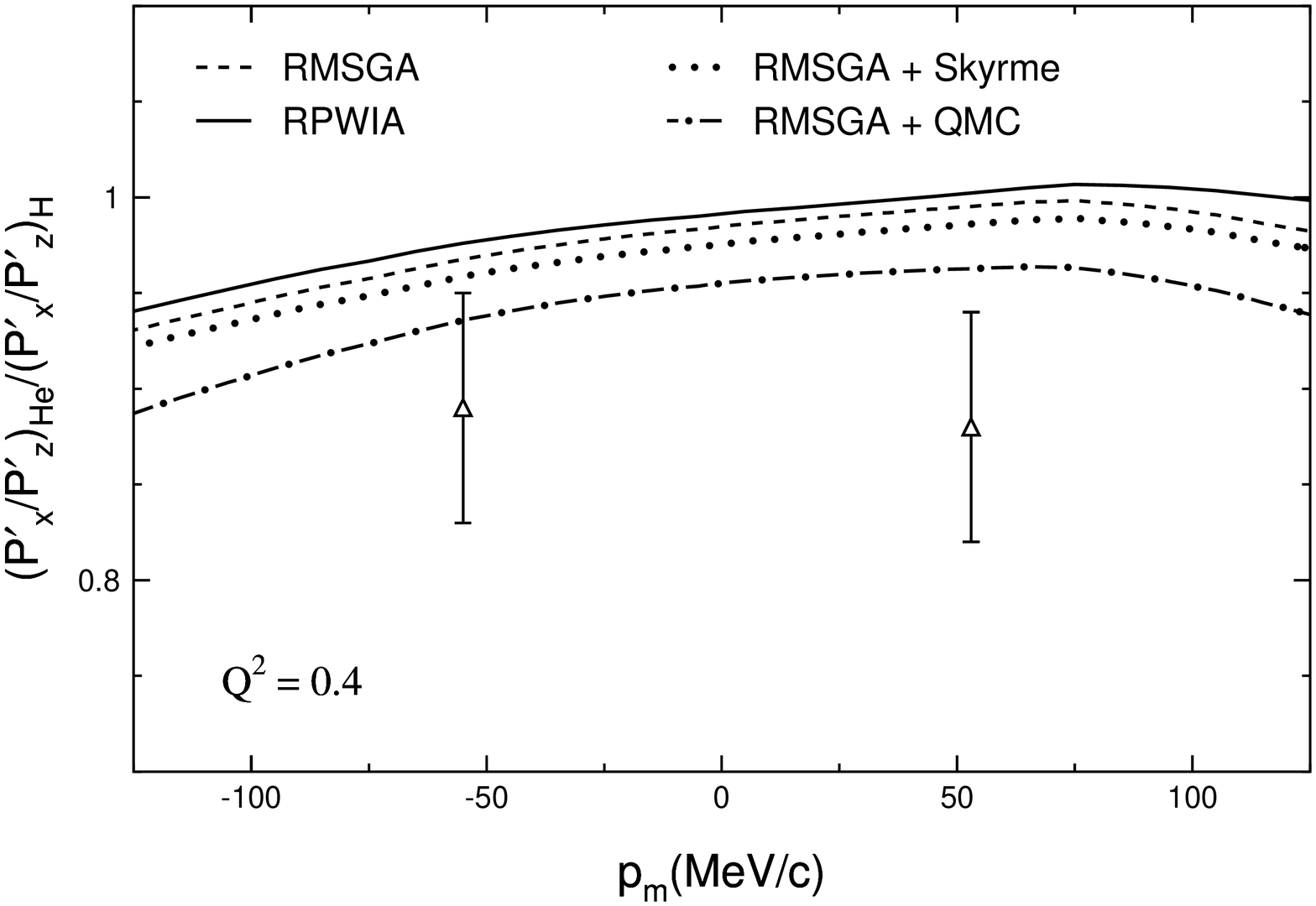,width=0.5\textwidth}}\quad
  \subfigure{\epsfig{figure=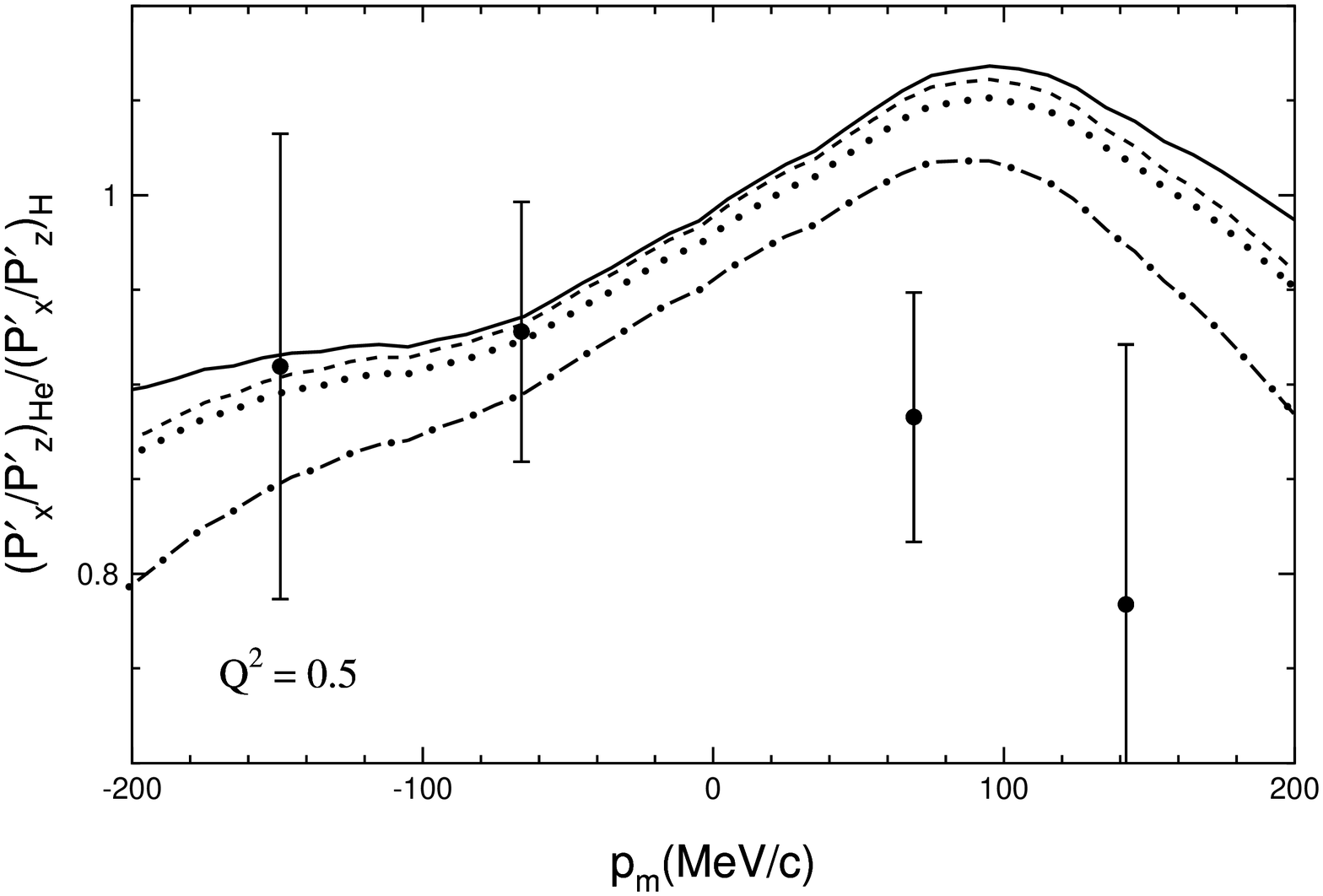,width=0.5\textwidth}}}
\end{center}
\caption{The double ratio $R$ as a function of the missing momentum
  for $Q^2 \mbox{=} 0.4$ and $0.5$~(GeV/c)$^2$ in $^4$He. The solid
  (dashed) curve are RPWIA (RMSGA) calculations. Influence of medium
  modifications are shown for the QMC (dot-dashed) and the Skyrme
  model (dotted). Data points are from \cite{dieterich01}(open
  triangles) and \cite{strauch03}(solid circles).  }
\label{fig.:dieterichsuperratio}
\end{figure*}

\begin{figure}
\begin{center}
\includegraphics[width=0.5\textwidth]{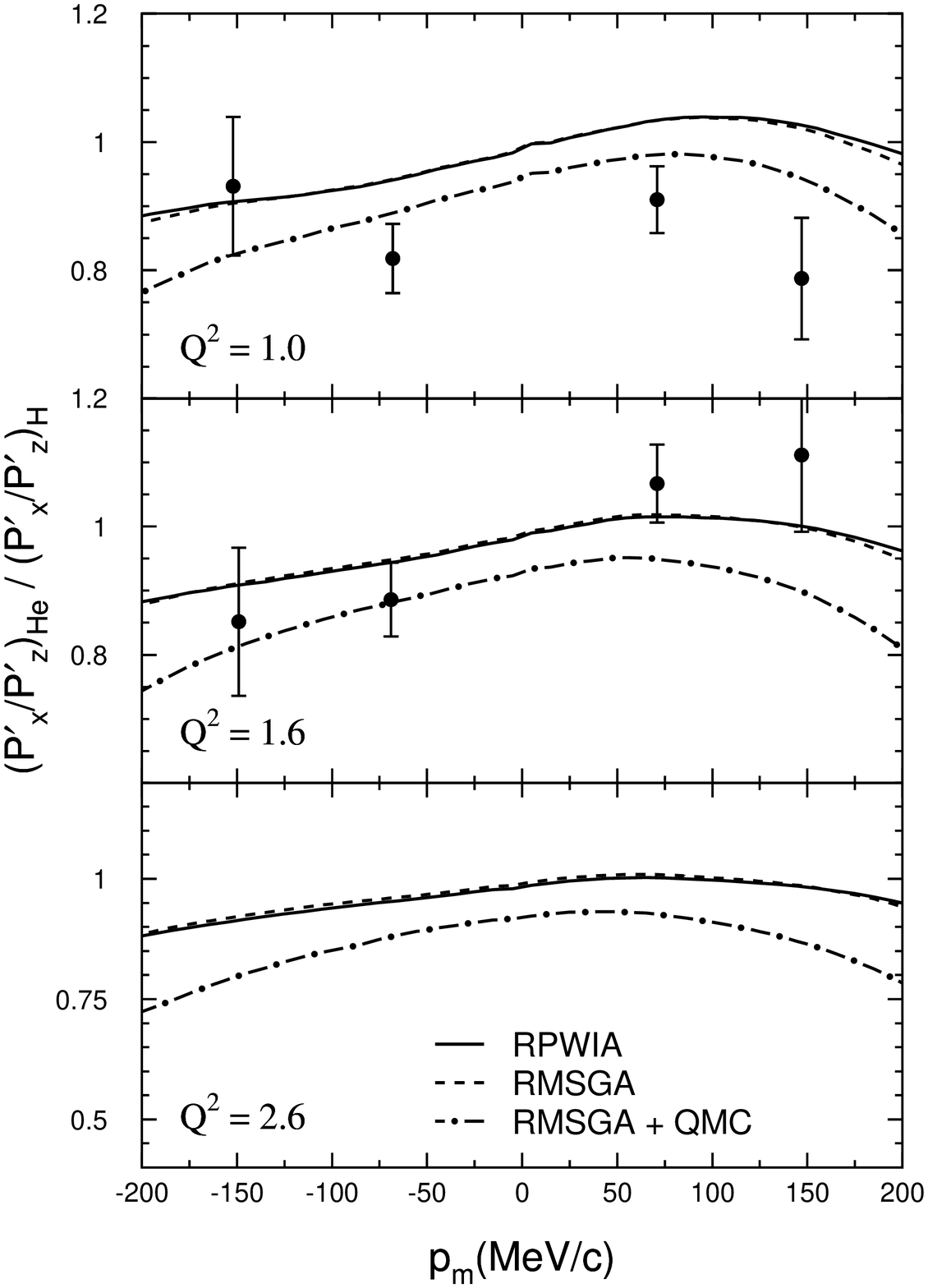}
\end{center}
\caption{The double ratio $R$ as a function of the missing momentum at
  three values of $Q^2$ in $^4$He. The solid (dashed) curve are RPWIA (RMSGA)
  calculations, while the dot-dashed curve represents RMSGA
  calculations including in-medium electromagnetic form factors of the
  QMC model. Data points are from \cite{strauch03}. }
\label{fig.:strauchsuperratio}
\end{figure}

A source of theoretical uncertainty on the computed
polarization-transfer components is the presence of short-range
correlations (SRC).  The RMSGA formalism outlined in
Sect.~\ref{sec:rmsga} is based on an independent-particle
approximation.  The effect of SRC on the FSI mechanisms can be
estimated by introducing a central correlation function in the
expression for the Dirac-Glauber phase of
Eq.~(\ref{eq.:glauberphase}).  This amounts to performing the
following substitution
\begin{equation}
  |\phi_{\alpha}(\vec{r}~')|^2 ~ \rightarrow ~
   |\phi_{\alpha}(\vec{r}~')|^2~ g(\vec{r} - \vec{r}~') \; ,
\label{eq:srcglaub}
\end{equation}
where $g(\vec{r} - \vec{r}~')$ is the central correlation function.
Physically, the existence of a central correlation function reflects
the inability of mean-field models to properly implement the strong
repulsion of the nucleon-nucleon force at short internucleon distances.
We use the central correlation
function from a G-matrix calculation by Gearheart and Dickhoff
\cite{gearheart94}. To date, the strongest sensitivity to central
correlation functions is observed in exclusive $A(e,e'pp)$
reactions. The adopted correlation function provides a favorable
agreement with the $^{12}$C$(e,e'pp)$ and $^{16}$O$(e,e'pp)$ data
\cite{ryckebusch04}. In the process of computing the Dirac-Glauber
phase of Eq.~(\ref{eq.:diracglauberphase}), the introduction of a
correlation function through the replacement of
Eq.~(\ref{eq:srcglaub}), strongly reduces the interaction between the
struck proton and any of the scattering centers when they are very
close (internucleon distances smaller than 0.8~fm) and bring about a
moderate enhancement for internucleon distances between 0.8 and 2
fm. In Fig.~\ref{fig.:src}, we investigate the effect of SRC on the
transferred-polarization components in $^4$He at two different values
of $Q^2$.  The results are expressed in the barycentric frame with $l$
parallel to the direction of the ejectile $\vec{p}_f$ and $t$ in the
hadronic plane, orthogonal to the $l$ component.  As we can see, the
SRC effects are relatively small, being typically of the order of 1\%
at a missing momentum of 200~MeV/c. Some asymmetric effect on $P'_l$
and $P'_t$ is seen.  A major finding is that the effect of SRC on
the Dirac-Glauber phase tends to cancel in the ratio $R$ at smaller
values of $Q^2$.  At higher values, we predict a modest enhancement of
$R$ due to SRC effects.

We now turn to the results for the double-polarization ratio $R$
obtained for the $^4$He nucleus. Response functions from the model
calculations were used in a Monte-Carlo code \cite{Ulmer} to calculate
the transferred and induced proton polarizations averaged over the
experimental acceptance. The starting point is always the huge number
of events (experimental data or MC simulations) within the acceptance
of the detectors. The full acceptance is then divided in various
bins. For Figs. \ref{fig.:dieterichsuperratio} and
\ref{fig.:strauchsuperratio} there are four bins in $p_m$ for the data
and several more for the calculations. Next, the average value of the
polarization is calculated for each bin. For the $p_m$
distributions the data is reported at the mean value of the missing
momentum within that bin. The best comparison with the model would be
to bin the MC data into the same number of bins as the data. One would
then compare one data point with one calculated point. That way,
however, the reader loses the information about the general
missing momentum dependence. Our comparison is reliable as long as the
transferrred polarizations are not changing rapidly within the
considered bin width.

Fig.~\ref{fig.:dieterichsuperratio} shows $R$
as a function of the missing momentum at $Q^2 \mbox{=}0.4$ and
$0.5$~(GeV/c)$^2$. We note that positive missing momentum $p_m$
corresponds to $|\vec{p}_f| < |\vec{q}|$. As can be inferred, the FSI
have only a minor impact on $R$, but
move the predictions somewhat closer to the measurements.  Both RMSGA
and RPWIA overestimate the double ratio $R$ by nearly $10\%$ and
predict $R \approx 1$ for zero recoil momentum.  After implementing
the medium-modified electromagnetic form factors from the QMC model, the
computed double-ratios $R$ are lowered by almost $8\%$, resulting in a
better overall agreement with the data.  The Skyrme model predicts
modest medium modifications which do not suffice to bring about a
major improvement in the description of the data within the context of
the RMSGA model.

Figure ~\ref{fig.:strauchsuperratio} summarizes the missing momentum
dependence of the $^4$He results for $Q^2 \ge 1~$ (GeV/c)$^2$
\cite{strauch03}. The FSI effects on $R$ are even smaller in this
high-energy regime. For $Q^2$=1.6~(GeV/c)$^2$ the measured $p_m$
dependence can be reasonably reproduced using free-nucleon form
factors.  Substituting the free form factors with the QMC ones reduces
$R$, an effect which grows with $p_m$. At $Q^2$=1.0~(GeV/c)$^2$ the
effect of medium modifications moves the theoretical curves
closer to the data. Qualitatively our RMSGA results are not
dramatically different from the RDWIA predictions presented in \cite{strauch03}.

\begin{figure}
  \begin{center}
    \includegraphics[width=0.5\textwidth]{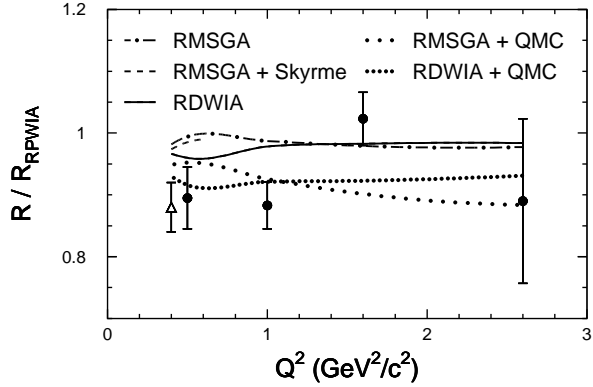}
  \end{center}
    \caption{The superratio $R/R_{RPWIA}$ as a function of $Q^2$ in
      $^4$He. The dot-dashed (solid) curve shows RMSGA (RDWIA)
      calculations, the dotted (dashed) curve represents RMSGA
      calculations with in-medium electromagnetic form factors from
      the QMC (Skyrme) model. The RDWIA and RDWIA+QMC results are
      those from the Madrid group as reported in
      Ref.~\cite{strauch03}. Data are from
      Refs.~\cite{dieterich01}(open triangle) and
      \cite{strauch03}(solid circles).}
    \label{fig.:doublesuperratio}
\end{figure}

In Fig.~\ref{fig.:doublesuperratio}, the superratio $R/R_{RPWIA}$ is
displayed as a function of $Q^2$.  Also here, the data and
calculations are integrated over the full experimental acceptance.
The data and calculations are reported as single points at the nominal
$Q^2$ value.  The model ``curves'' only connect the computed points to
guide the eye.  As seen in Fig.~\ref{fig.:doublesuperratio} the
Mainz data point nicely matches with the lowest $Q^2$ measurement at
JLAB. As off-shell effects are not completely negligible for the
polarization-transfer components, it is worth stressing that the RDWIA
(RMSGA) $^4$He results shown here are obtained with the $CC1$ ($CC2$)
current operator.  For $Q^2 \le$1~(GeV/c)$^2$ the standard nuclear
physics RDWIA and RMSGA results fail to reproduce the ratio $R$.  The
overestimation is of the order of 10\% for RMSGA, and 5-7\% in
RDWIA. The predicted four-momentum dependence for $R$ is modest in
both models. The RMSGA attributes smaller effects to FSI than RDWIA
does. In Ref.~\cite{lava03}, a similar trend was found when comparing
RDWIA and RMSGA $A(e,e'p)$ nuclear transparencies for light nuclei.

\begin{figure}
\begin{center}
\includegraphics[width=0.5\textwidth]{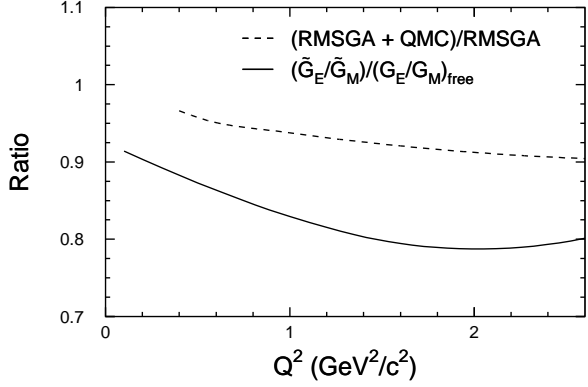}
\end{center}
\caption{The ratio of the RMSGA+QMC to the RMSGA prediction for $R$ as
  a function of $Q^2$ for the $1s1/2$ proton in $^4$He (dashed
  line). The solid line shows $\biggl[ \widetilde{G}_E^{QMC} (1s1/2,Q^2)
  /\widetilde{G}_M^{QMC} (1s1/2,Q^2) \biggr] / \biggl[ G_E (Q^2) / G_M (Q^2)
  \biggr] $.}
\label{fig.:gintratio}
\end{figure}

Inclusion of medium modifications for the electromagnetic form factors
according to the predictions of the Skyrme model shifts the RMSGA
calculations marginally closer to the data. The results for the Skyrme
model are shown up to $Q^2$=0.6~(GeV/c)$^2$ since the model is no
longer deemed realistic at higher values.  Implementing QMC
electromagnetic form factors, on the other hand, lowers the
$p_m$-integrated RMSGA predictions for the superratio $R$ between 5\%
and 10\%. The difference between the RMSGA and the RMSGA+QMC values
for $P'_x/P'_z$ grows with increasing $Q^2$.  This reflects the fact
that in the QMC model, the ratio $\widetilde{G}_{E}/\widetilde{G}_{M}$
moves steadily away from the free values with increasing $Q^2$ to
reach a maximum of over 20\% at about $Q^2$=2~(GeV/c)$^2$, after which
some turning in the trend is observed.  This has been illustrated in
Fig.~\ref{fig.:gintratio}. As can be inferred from this picture, about
one third of the predicted magnitude of the medium modifications on
$G_{E}/G_{M}$ is visible in the $P'_x/P'_z$ ratio.  It is worth
stressing that Fig.~\ref{fig.:gintratio} compares two different
quantities.  On the one hand, the curve showing the
$\widetilde{G}_{E}/\widetilde{G}_{M}$ has been averaged over the
squared $1s1/2$ proton overlap wave function, thus receiving its
largest contributions from the nuclear interior.  This is not
necessarily the case for the $^{4}$He$(\vec{e},e'\vec{p})$
observables.  Indeed, in the process of computing the
observables, the medium effects in the form factors are weighted with
a more complex function which involves not only the $1s1/2$
proton overlap wave function, but also the current operator and the
scattering wave function. The dashed curve of Fig.~\ref{fig.:gintratio} indicates
that over the whole larger radii, and correspondingly lower densities,
are probed.  This phenomenon reduces the magnitude of the
medium-dependent effects on the observables.

\begin{figure}
\begin{center}
\includegraphics[width=0.5\textwidth]{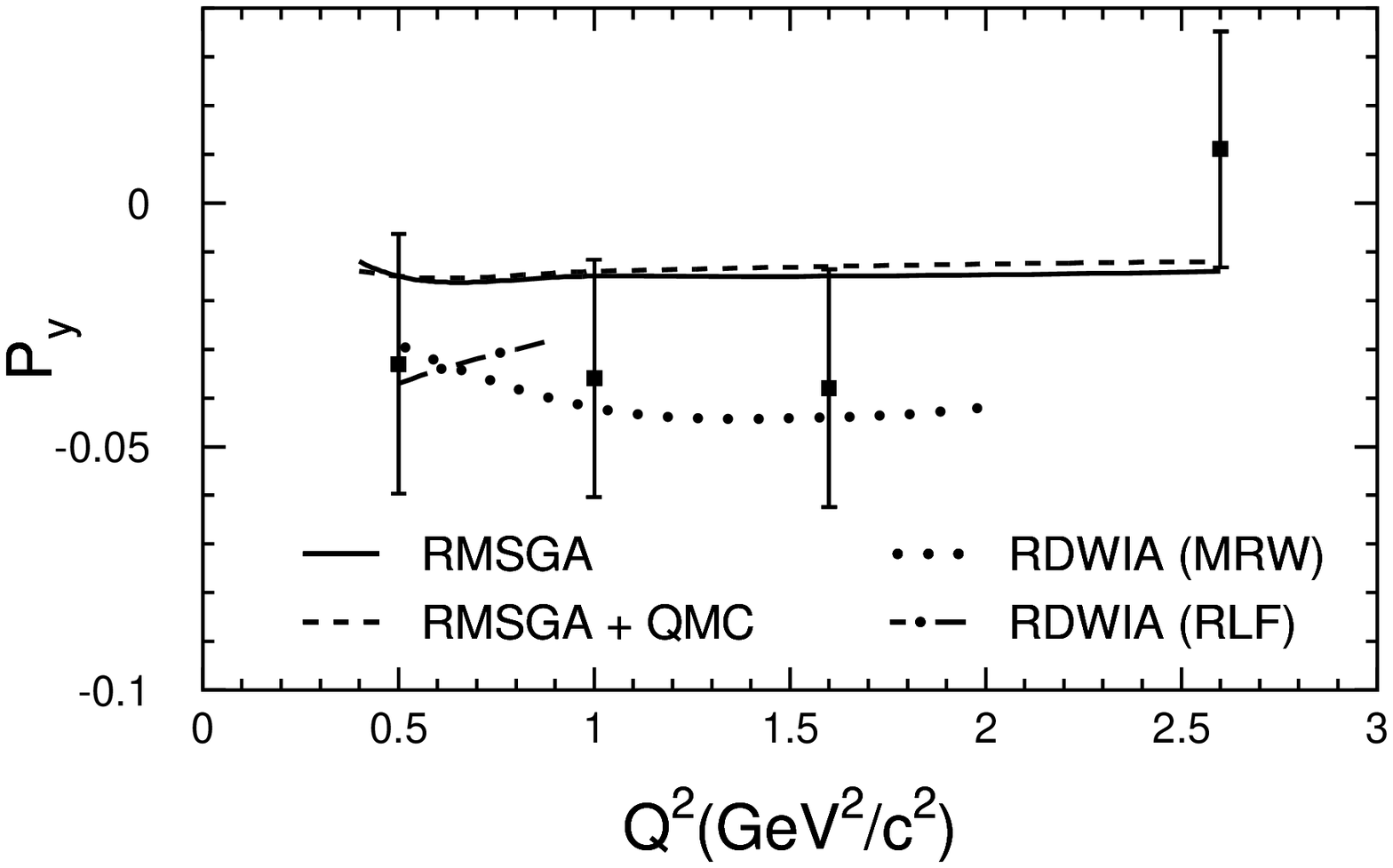}
\end{center}
\caption{ The induced normal polarization as a function of $Q^2$ in
  $^4$He. The solid curve represents RMSGA calculations with free from
  factors. For the dashed lines $\widetilde{G}_{E,M}^{QMC}(1s1/2,Q^2)$
  form factors are used. Data points and RDWIA results are from
  Ref.~\cite{strauch03}.}
\label{fig.:strauchpn}
\end{figure}

Figure~\ref{fig.:strauchpn} displays the induced normal polarization
as a function of $Q^2$ for the $^4$He nucleus. In the one-photon
exchange approximation, $P_y$ vanishes in the absence of
FSI. Accordingly, this observable serves as a stringent test for
models of FSI mechanisms. The smallness of $P_y$ suggests relatively
moderate FSI. The RDWIA calculations for $P_y$ are shown for exactly
the same kinematics, though with the $CC1$ choice for the current
operator.  The RDWIA predictons for the $P_y$ in $^4$He$(e,e'\vec{p})$
are presented for two viable choices of the optical-potential
parameterization~: ``RLF'' (limited to proton lab kinetic energies
smaller than 0.4~GeV) and ``MRW'' (limited to proton lab kinetic
energies smaller than 1.0~GeV).  The two optical potentials predict a
dissimilar $Q^2$ dependence for $P_y$.  Indeed, in many cases various
optical potentials can fit the elastic proton-nucleus data equally
well, but do not necessarily lead to identical predictions in
electromagnetically induced nucleon knockout. The RDWIA model predicts
values for $P_y$ which are over twice as large than the RMSGA ones.
Studies in the Dirac eikonal approach have stressed the importance of
the spin-orbit part in the optical potential for the computed values
of $P_y$ in $^{16}$O \cite{ito97}. Similar observations have been made
for the $^{12}$C$(e,e'\vec{p})$ case at $Q^2 \mbox{=}$
0.49~(GeV/c)$^2$ \cite{woo98}. The measured value of $P_y$ at
$Q^2$=2.6~(GeV/c)$^2$ may indeed suggest the decreasing role of this
spin-dependent part as the energy increases.  As can be inferred,
$P_y$ remains nearly unaffected by medium modifications in the
electromagnetic form factors.  This is not unexpected given that $P_y$
is an observable which quantifies the magnitude of secondary
processes, like rescattering mechanisms.  The introduction of
medium-modified form factors induces some change in the way these
mechanisms are folded over the density of the target nucleus.

\begin{figure*}
\begin{center}
\mbox{\subfigure{\epsfig{figure=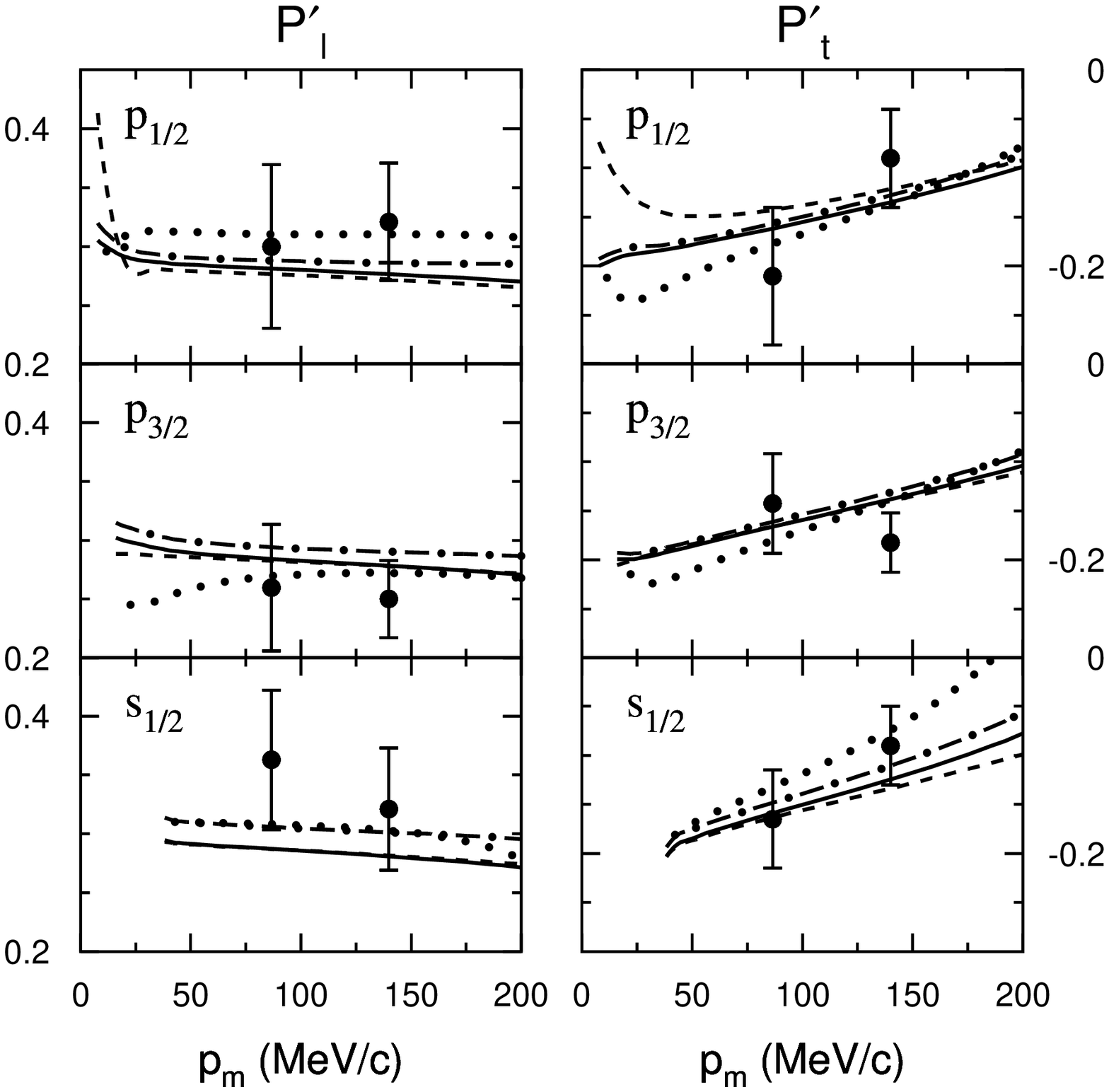,width=0.5\textwidth}}\quad
  \subfigure{\epsfig{figure=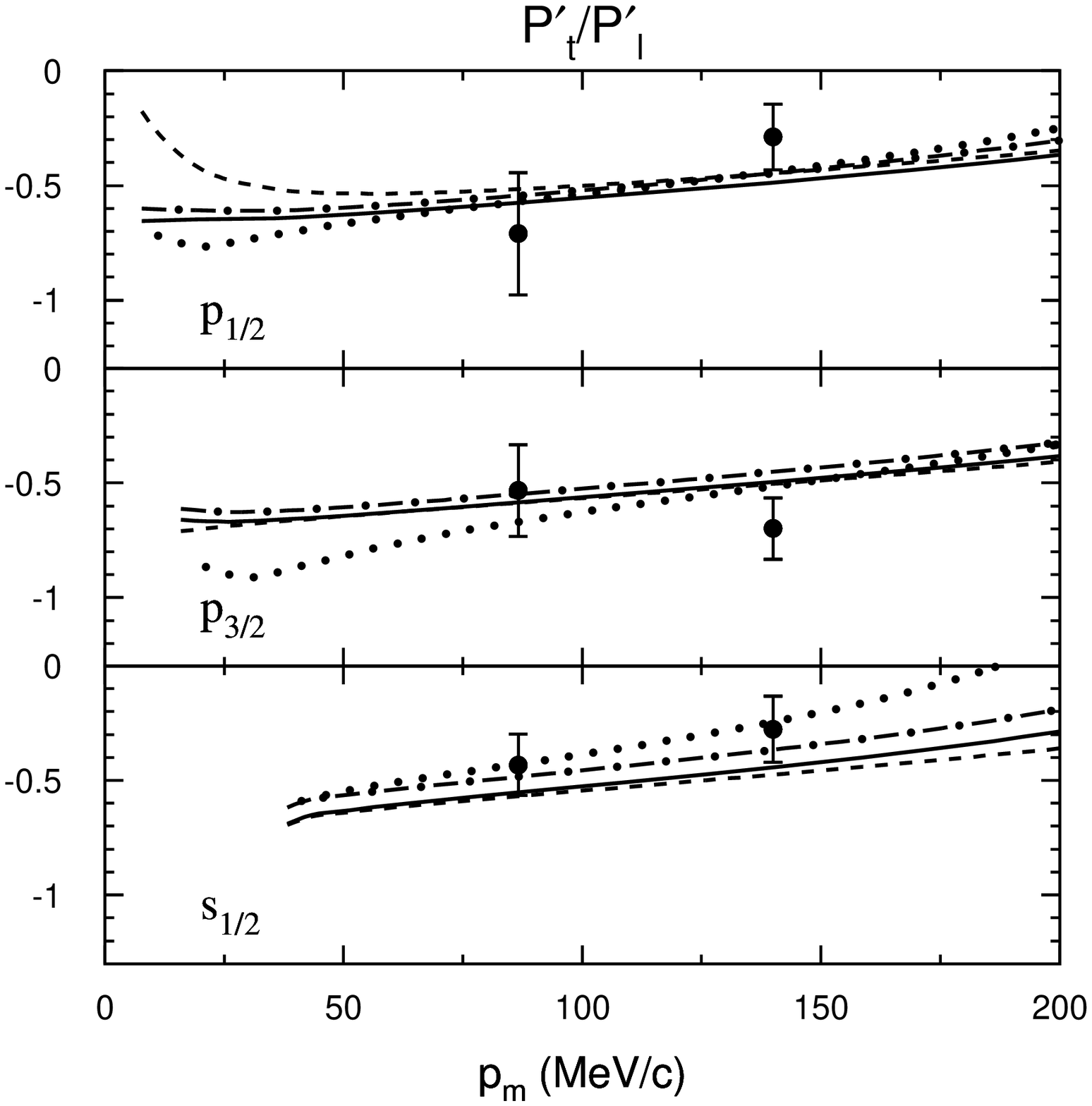,width=0.5\textwidth}}}
\end{center}
\caption{ Transferred polarization components for proton knock-out
  from the three shells in $^{16}$O for $Q^2 \mbox{=} 0.8$ (GeV/c)$^2$
  in constant $(\vec{q},\omega)$ kinematics. The solid (dashed) curve
  represents RMSGA (RPWIA) calculations with free-nucleon form
  factors, while the dot-dashed curve is obtained from RMSGA
  calculations when using the QMC form factors. The dotted curve
  represents RDWIA calculations. Data points are from \cite{malov00}.}
\label{fig.:malov}
\end{figure*}

Finally, in Fig.~\ref{fig.:malov}, results for the transferred
polarization components and their ratio for the $^{16}$O nucleus are
shown at $Q^2 \mbox{=}0.8$ (GeV/c)$^2$.  Hereby, we adopt constant
($\vec{q}, \omega$) kinematics and compare the RMSGA predictions with
the measurements of Ref.~\cite{malov00} and the results of the RDWIA
model from the Madrid group.  For the oxygen calculations, the RDWIA
and RMSGA calculations adopt identical mean-field wave functions (W1
parameterization) and current operators ($CC2$ in the Coulomb
gauge). The RDWIA calculations are performed with the EDAD1
parameterization for the optical potential \cite{cristina}.  In
essence, the RDWIA and RMSGA curves result from identical model
ingredients, apart from the implementation of FSI distortions which is
grounded on different philosophies. The RPWIA and RMSGA curves for
$P'_l$ and $P'_t$ are close, the RDWIA model predicting larger FSI
distortions.  At corresponding $Q^2$ values
(Figs.~\ref{fig.:dieterichsuperratio} and
\ref{fig.:strauchsuperratio}), the $^4$He results could be better
reproduced after introducing QMC medium-modified form factors.  As can
be appreciated from Fig.~\ref{fig.:malov}, the $^{16}$O data do not
allow one to draw conclusions on the possibility of medium
modifications.  The overall trends of the $^{16}$O
polarization-transfer data are reasonably reproduced in the RMSGA,
using free-proton electromagnetic form factors.  When comparing the
RMSGA and RMSGA+QMC curves a significant orbital dependence of the
magnitude of the medium effect is observed. This orbital dependence
can be partly attributed to the use of a weight function depending on
$\rho _{\alpha_1} (\vec{r})$ in Eq.~(\ref{eq.:currentoperatordens}). Comparing
the results for $R$ for the various orbitals in a particular nucleus
could allow one to study the density dependence of the medium
effects.

\section{Conclusions}
\label{sec:conc}

Unfactorized and relativistic Glauber calculations are performed for
the polarization-transfer components in $^4$He and
$^{16}$O$(\vec{e},e'\vec{p})$ for $Q^2$=0.4, 0.5, 0.8, 1.0, 1.6 and
2.6~(GeV/c)$^2$.  The selected kinematics are those for which data are
available. The adopted framework has the virtues that it is
relativistic and can be reliably applied up to the highest
four-momentum transfers covered in the measurements.  Overall, the
effect of FSI on the polarization-transfer components is smaller in
the relativistic Glauber framework than in a relativistic
optical-potential framework.  After all, this is not so surprising
given that typical Glauber approaches rely on spin-independent
nucleon-nucleon scattering amplitudes when modeling the final-state
interactions.  The spin-dependent effects are expected to lose their
importance as the energy increases.  Polarization studies with the
electromagnetic probe, like the one presented here, will help in
further clarifying this issue.

For the $^{16}$O target, for which the data are restricted to
$Q^2$=0.8~(GeV/c)$^2$, the calculations provide a fair description
when adopting free-proton electromagnetic form factors.  A similar
situation holds for the $^4$He case at $Q^2 \ge $1.6~(GeV/c)$^2$.  For
$^4$He and $Q^2 \le $1.0~(GeV/c)$^2$ substantial deviations between
the RMSGA predictions and the data are observed.  Under these
circumstances, the implementation of the in-medium form factors from
the QMC nucleon model, makes the RMSGA calculations to go in the right
direction and induces changes in the ratio of the polarization-transfer
components, which are of the right order of magnitude to explain the
discrepancies.  A recently approved experiment at JLAB \cite{ent} will
address the polarization-transfer ratio at $Q^2$-values of $0.8$ and
$1.3$~(GeV/c)$^2$, and is expected to reduce the statistical
uncertainties by over a factor of two compared to the previous round
of measurements.

\section{Acknowledgment}
We wish to thank J.M. Ud\'{i}as and M.C. Mart\'{i}nez for providing us
with the RDWIA calculations and for numerous discussions. This work is
supported by the Fund for Scientific Research (FWO), the research
council of Ghent University and by the U.S. Department of Energy under
grant DE-FG02-95ER40901.


\end{document}